\documentclass{aa}

\newcommand{\kepler}{{\it Kepler}}

\newcommand{\tess}{TESS}

\begin{document}

\title{Searching for the mHz variability in the \tess\ observations of nova-like cataclysmic variables}
\titlerunning{mHz variability in nova-likes}
\authorrunning{Dobrotka et al.}

\author{A.~Dobrotka \inst {1}, J.~Magdolen \inst {1}, and D.~Jan\'ikov\'a \inst {1}}

\institute{Advanced Technologies Research Institute, Faculty of Materials Science and Technology in Trnava, Slovak University of Technology in Bratislava, Bottova 25, 917 24 Trnava, Slovakia
}

\date{Received / Accepted}

\abstract
{}
{We investigated fast optical variability of selected nova-like cataclysmic variables observed by \tess\ satellite. We searched for break frequencies ($f_{\rm b}$) in the corresponding power density spectra (PDS). The goal is to study whether these systems in almost permanent high optical state exhibit preferred $f_{\rm b}$ around 1\,mHz.}
{We selected non-interrupted light curve portions with duration of 5 and 10 days. We divided these portions into ten equally long light curve subsamples and calculated mean PDS. We searched for $f_{\rm b}$ in frequency interval from log($f$/Hz) = -3.5 to -2.4. We defined as positive detection when the $f_{\rm b}$ was present in at least 50\% of the light curve portions with a predefined minimum number of detections.}
{We measured $f_{\rm b}$ in 15 nova-like systems and confirmed that the value of this frequency is clustered around 1\,mHz with a maximum of the distribution between log($f$/Hz) = -2.95 and -2.84. The confidence that this maximum is not a random feature of a uniform distribution is at least 96\%. This is considerably improved since previous value of 69\%. We discuss the origin of these $f_{\rm b}$ in the context of sandwich model where central hot X-ray corona surrounds central optically thick disc. This scenario could be supported by correlation between white dwarf mass and $f_{\rm b}$; the larger the mass, the lower the frequency. We see such tendency in the measured data, however the data are too scattered and based on low number of measurements. Finally, it appears that systems with detected $f_{\rm b}$ have lower inclination than 60-75$^{\circ}$. In higher inclination binaries the central disc is not seen and the PDS is dominated by red noise. This also supports the inner disc regions as source of the observed $f_{\rm b}$.}
{}

\keywords{accretion, accretion discs - stars: novae, cataclysmic variables - stars: white dwarfs}

\maketitle

\section{Introduction}
\label{introduction}

Cataclysmic variables (CVs) are interacting binaries powered by an accretion process. The mass is transferred from a main sequence companion star via Roche lobe overflow, and in the absence of a strong magnetic field, an accretion disc forms (see e.g. \citealt{warner1995} for a review). In any case, the matter flows towards the central white dwarf (WD).

In CVs we distinguish between high and low optical brightness states based on the mass accretion rate $\dot{m}_{\rm acc}$ through the disc. The physical conditions in these states are defined by hydrogen ionisation, and the alternation between these two states observed mainly in dwarf novae is generated by viscous-thermal instability (\citealt{osaki1974,hoshi1979,meyer1981}). In the low state the disc is cold, the hydrogen is recombined and $\dot{m}_{\rm acc}$ is low. During the high state, the hydrogen is ionised in the hot disc and is brighter in the optical due to the larger $\dot{m}_{\rm acc}$ compared to the low state. The accretion disc is fully developed (almost) up to the WD in the high state, while it is truncated in the low state. This truncation explains delays between optical, UV radiation and X-rays (see e.g. \citealt{schreiber2003}).

CVs with mass transfer from the secondary high enough, spend most of their lifetime in a high state. This is due to $\dot{m}_{\rm acc}$ being above a critical value, and the viscous-thermal instability does not appear. This larger matter flow through the disc ensures an almost permanently hot disc with ionised hydrogen and dwarf nova outbursts suppression. Such CVs systems are called nova-likes. Fluctuations in mass transfer from the secondary, probably due to star spots, occasionally appear, and the nova-like system falls to a low state for a relatively short time (\citealt{honeycutt2004}). Such behaviour is described as antidwarf novae and is typical for VY\,Scl nova-like systems.

Typical manifestation of the underlying accretion process is fast stochastic variability called flickering. This variability exhibits variety of observational features with the most important being: (1) linear correlation between variability amplitude and log-normally distributed flux (so called rms-flux relation) (see \citealt{scaringi2012b,vandesande2015}), (2) power density spectra (PDS) with the shape of a red noise or band limited noise with characteristic frequencies in the form of a break or Lorentzian (\citealt{scaringi2012a,dobrotka2016}) and (3) time lags, in which the flares reach their maxima slightly earlier in the blue than in the red (\citealt{scaringi2013,bruch2015}).

The characteristic frequencies in the PDSs bring information about time scales of physical processes generating the variability. Some systems show a single break frequency $f_{\rm b}$ in the PDS like UU\,Aqr (\citealt{baptista2008}) or KR\,Aur (\citealt{kato2002}). Usually, these detections are made using ground observations. For detection of multicomponent PDS a long and uninterrupted light curve is needed. This is the case of \kepler\ spacecraft, which allowed to detect various components in MV\,Lyr (\citealt{scaringi2012a}), V1504\,Cyg (\citealt{dobrotka2015}) and V344\,Lyr (\citealt{dobrotka2016}). X-ray observations yielded also detections of single or multicomponent PDSs. Examples of single-component PDSs are VW\,Hyi, WW\,Cet, T\,Leo (\citealt{balman2012}), while SS\,Cyg (\citealt{balman2012}), RU\,Peg (\citealt{dobrotka2014}) or MV\,Lyr (\citealt{dobrotka2017}) exhibit more complex PDS morphology.

\citet{dobrotka2020} summarized various detections of characteristic frequencies of flickering in the PDSs of CVs. The study implies that two characteristic frequencies can exist in the low state. The same is seen in the high state, but together with an additional third most prominent component at log($f$/Hz) $\simeq$ -3. However, the study summarized only 12 detections, and the corresponding frequency distribution can be easily explained by the random appearance of a uniform distribution. More detections are needed to construct a significant histogram.

\citet{dobrotka2021} found another system exhibiting the log($f$/Hz) $\simeq$ -3 signal as a clear $f_{\rm b}$ in \kepler\ data. However, the system is an intermediate polar (IP) V4743\,Sgr with a magnetic WD. Such magnetic CVs have truncated discs which makes them different from non-magnetic nova-like CVs. $f_{\rm b}$ with values close to or higher than log($f$/Hz) $\simeq$ -3 were detected in other IPs (\citealt{revnivtsev2010,semena2014}). The detected $f_{\rm b}$ are very close to WD spin values. If the inner truncated disc edge is co-rotating with the central WD, this explains the similarity of the break and spin frequency. Therefore, such $f_{\rm b}$ at log($f$/Hz) $\simeq$ -3 is non-associated to the flickering activity in CVs. However, V4743\,Sgr is rather special IP. It is a nova remnant where the mass transfer rate is quite high, and the inner disc edge should be pushed much closer to the WD, and further away from the co-rotating radius. Therefore, the detected $f_{\rm b}$ can still be associated to the flickering activity, even if its value is close to the WD spin.

In order to test the hypothesis about preferred characteristic frequencies in CVs in high and low state, new PDSs studies are needed. The probability that the characteristic frequency log($f$/Hz) $\simeq$ -3 is preferred is only 69\% so far (\citealt{dobrotka2021}). With a questionable magnetic V4743\,Sgr the probability rises to 91\%.

\citet{scaringi2014} interpreted the main feature at log($f$/Hz) $\simeq$ -3 detected in optical \kepler\ data of MV\,Lyr as being due to a hot, geometrically thick disc (hot X-ray corona) surrounding a cool, geometrically thin disc (the so called sandwich model). Such corona consists of evaporated gas from the underlying geometrically thin disc (\citealt{meyer1994}). It radiates in X-rays, and these X-rays are reprocessed into optical by the geometrically thin disc. The physical origin of the variability is explained by propagating mass accretion fluctuations (\citealt{lyubarskii1997,kotov2001,arevalo2006}) in the corona.

However, the ratio of X-ray to optical luminosity is on the order of 0.1 (see e.g. \citealt{dobrotka2020}), which is too low to explain the observed optical and UV variability with a reprocessing scenario. \citet{dobrotka2019} studied the shot profile of the flickering in the \kepler\ data of MV\,Lyr. The authors found two components with different amplitudes in the averaged shot profile. Both having characteristic frequencies of log($f$/Hz) $\simeq$ -3. While the component with high amplitude (central spike) is problematic to be generated by the reprocessing, the low amplitude side-lobes match the scenario (\citealt{dobrotka2020}). This suggests that the variability is formed in two separate regions: the geometrically thin disc, and the reprocessed X-rays of the geometrically thick corona.

Apparently, the origin of the flickering variability is not yet understood, and further study is needed. Our goal is to increase the number of measured PDS characteristic frequencies. The corresponding distribution of frequency values can have substantial impact on flickering understanding, like it was in the case of orbital period distribution (see \citealt{warner1995} for review). In this paper we searched for $f_{\rm b}$ in PDSs calculated from \tess\ data of nova-like CVs.

\section{Selected systems and observations}

In this paper we present analysis of \tess\ light curves of all nova-like and old nova systems listed in the \citet{bruch2022}, \citet{bruch2023a} and \citet{bruch2023b}. The Pre-search Data Conditioning Simple Aperture Photometry (PDCSAP) light curves were downloaded\footnote{Using Python Lightkurve library; \url{https://docs.lightkurve.org}.} from the Mikulski Archive for Space Telescopes (MAST) with cadence of 120 seconds as with PDCSAP flux long term trends are removed.

\section{PDS analysis}

We searched for characteristic $f_{\rm b}$ in PDSs calculated from the selected light curves. Since \kepler\ spacecraft is much superior instrument, the corresponding PDSs showed multicomponent character in several cases (\citealt{scaringi2012a,dobrotka2016}). The \tess\ case is considerably worse, and only the dominant $f_{\rm b}$ can be seen. Moreover, selected nova-like CVs show superhump or orbital periods longer than 0.06\,day (\citealt{bruch2022,bruch2023a,bruch2023b}). This corresponds to frequency of log($f$/Hz) = -3.7 and lower. Such frequency can affect the observed PDS and lead to wrong identification or measurement of $f_{\rm b}$. However, we are interested mainly in high frequency component close to log($f$/Hz) $\simeq$ -3. Therefore selecting frequency interval from log($f$/Hz) = -3.5 to higher frequencies resolve the potential superhump or orbital period contamination problem.

\subsection{Method}

\begin{figure}
\resizebox{\hsize}{!}{\includegraphics[angle=0]{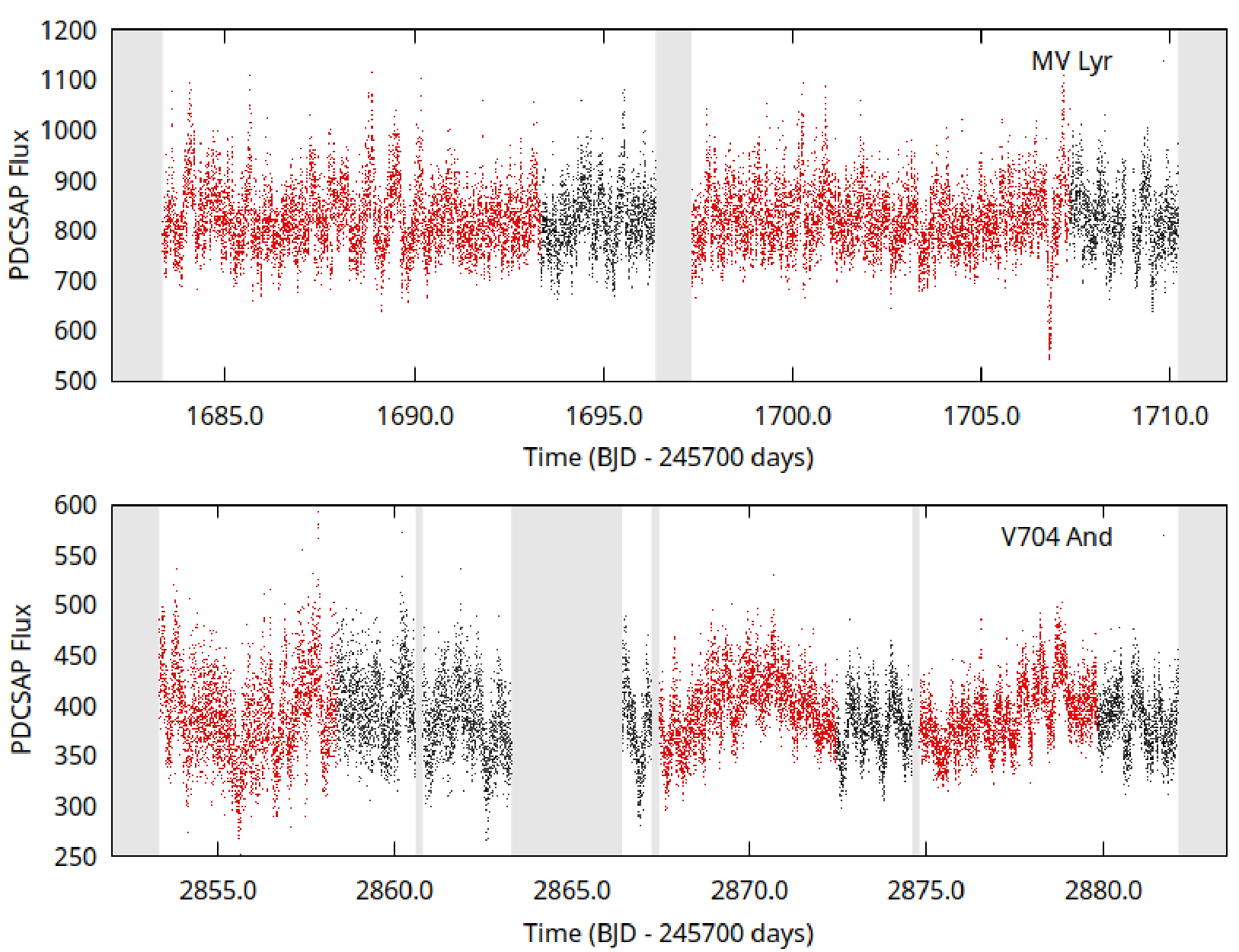}}
\caption{Examples of light curves (black points) with gaps (gray shaded areas) and selected portions (red points) with duration of 10 (upper panel) and 5 (lower panel) days for PDS analysis.}
\label{lc_examples}
\end{figure}

We first selected suitable light curve portions for PDS calculation. To avoid large gaps in the data, we selected only the segments between them. We defined a gap as a duration of 0.2 days or more. This limit was determined empirically, as using this threshold all obviously large gaps were eliminated. Fig.~\ref{lc_examples} shows two examples of this selection process.

Second step is to select the light curve portions with duration of $n_{\rm d}$ days. Empirically derived optimal value is 10. However, not all of the examined objects have such uninterrupted 10 days intervals. Therefore, we performed analysis using $n_{\rm d} = 5$ and 10. We could use the whole light curve portion between the gaps, but we prefer to have all PDSs equivalent with the same frequency resolution and PDS point step.

To estimate the PDS we divided each light curve portion into $n_{\rm subs}$ subsamples. For each subsample, we calculated a periodogram using the Lomb-Scargle algorithm\footnote{We used python's package {\tt Astropy} (\citealt{astropy_collaboration2013,astropy_collaboration2018,astropy_collaboration2022}).} (\citealt{scargle1982}). All periodograms (power\footnote{We use power normalized by the total variance according to \citet{horne1986}. Since the normalization does not affect the shape of the PDS, this has no importance in our case.} $p$ as a function of frequency $f$) were subsequently transformed into log-log space. Averaging of log($p$) instead of $p$ is recommended by \citet{papadakis1993}. The whole log($f$) interval of the PDS we re-binned with a constant frequency step of 0.1\,dex. Finally, all log($p$) points within each frequency bin were averaged. We defined an additional condition for averaging; the number of averaged periodogram points per bin must be larger than some pre-selected value defined as $n_{\rm min} \times n_{\rm subs}$ ($n_{\rm min}$ points from each of $n_{\rm subs}$ periodograms). If the condition is not valid, the bin should be broadened until the condition is fulfilled. This allows us to get enough points for mean value with standard error of the mean calculation.

The PDS frequency resolution and the lowest PDS frequency (before re-binning) are proportional to the duration of the light curve subsamples. Therefore, an empirical compromise between noise and resolution must be found. As $n_{\rm subs}$ and $n_{\rm min} \times n_{\rm subs}$ values we chose 10 and $3 \times 10$, respectively. The high-frequency end of the periodogram is set by the Nyquist frequency.

We fitted the resulting PDSs with a broken power law model using the {\small GNUPLOT}\footnote{\url{http://www.gnuplot.info/}} software. The selected frequency range was from log($f$/Hz) = -3.5 to -2.4.

\subsection{Selection of positive detections}

\begin{figure*}
\resizebox{\hsize}{!}{\includegraphics[angle=-90]{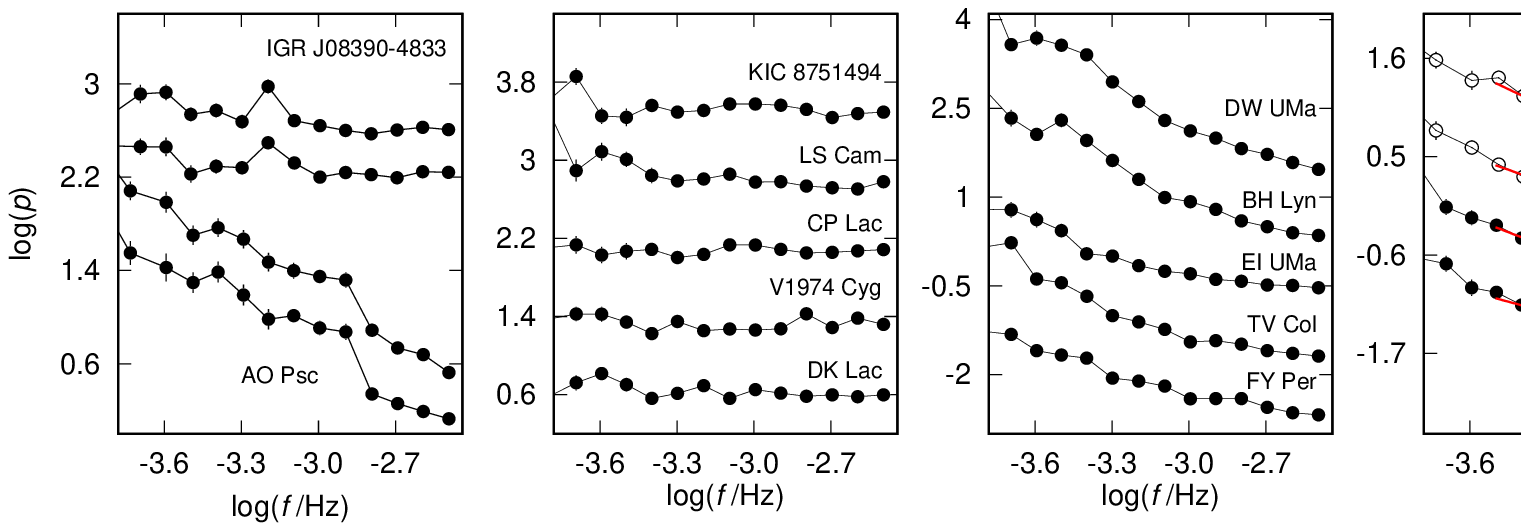}}
\caption{Examples of PDSs. Black points represent PDSs with $n_{\rm d} = 10$ and black circles are for $n_{\rm d} = 5$. First (left) panel - IPs with strong spin frequency affecting the PDS. Second panel - example of PDSs dominated by white noise. Third panel - example of PDSs dominated by red noise. Fourth (right) panel - Examples of PDSs of MV\,Lyr with broken power law fits. The upper PDS shows case where the fitting process did not converge to the broken power law shape. There is no $f_{\rm b}$ but a straight line instead. The middle two cases represent possible positive detection but with $f_{\rm b}$ error larger than 0.10. Apparently, the $f_{\rm b}$ is not clear even if the fit converged. The lower case is an example of a positive detection with a small enough error of the $f_{\rm b}$ due to clear broken power law shape of the PDS. Individual PDSs are offset vertically for better visualisation.}
\label{pdss_bad}
\end{figure*}

First step is to identify systems with PDS contaminated by other signals not associated to flickering. Superhumps and orbital periods are no longer a complication, as described above. However, IPs like AO\,Psc and IGR\,J08390-4833 have spin frequencies of log($f$/Hz) = -2.91 (\citealt{vanwderwoerd1984}) and -3.16 (\citealt{sazonov2008}), respectively. This clearly affects the studied PDSs (left panel in Fig.~\ref{pdss_bad}). These systems we excluded from the analysis.

Second step is to select objects with positive $f_{\rm b}$ detections. Systems, where the PDS is dominated by white (second panel of Fig.~\ref{pdss_bad}) or red noise (third panel of Fig.~\ref{pdss_bad}) did not yield any detection. More problematic are the cases, where random features or noise in the PDS can mimic the searched $f_{\rm b}$. For this reason we selected only those PDSs, where $f_{\rm b}$ is persistent and stable. This means, that it must be detected various times and with similar values. For this selection process we relied on well studied MV\,Lyr (\citealt{scaringi2012a}). The detection of $f_{\rm b}$ close to log($f$/Hz) = -3 is unambiguous thanks to the \kepler\ observations. After inspecting individual PDSs from the \tess\ light curves, we selected only those light curve subsamples where the detected $f_{\rm b}$ has an error of 0.10 or less\footnote{Since this error limit is more or less approximate and empirically estimated, we took all values rounded to 0.10, therefore up to 0.10499... This allowed us to get five additional measurements. This was useful mainly in AH\,Men case where it helped to fulfill the $n_{\rm p} =$ 50\% condition.}. Larger errors occurred in cases of clearly scattered and noisy PDSs (right panel in Fig.~\ref{pdss_bad}) or where the break was not obvious, making the broken power law shape is untrustworthy. Moreover, if the PDSs did not show the broken power law shape at all, the fit did not converge to such a shape, and the $f_{\rm b}$ reached the end with absurd errors. This selection process yield 19 detections of $f_{\rm b}$ out of 27 light curve subsamples with $n_{\rm d} = 5$, and 10 detections out of 13 light curve subsamples with $n_{\rm d} = 10$. Therefore, the fraction of positive detections $n_{\rm p}$ is 70\% and 77\% for $n_{\rm d} = 5$ and $n_{\rm d} = 10$, respectively. Apparently, the persistent $f_{\rm b}$ in MV\,Lyr is not detected in all light curve subsamples probably due to limited capabilities of \tess\ yielding more scattered PDSs compared to the \kepler\ results. Therefore, as positive detections in all other systems we choose cases where $n_{\rm p}$ parameter was at least 50\%. However, if the total number of light curve subsamples is only two, the detection of a single $f_{\rm b}$ is not sufficient. Therefore, we considered a detection to be positive if at least two $f_{\rm b}$ were detected. More strict condition will be applied later.

\subsection{Results}

\begin{figure*}
\resizebox{\hsize}{!}{\includegraphics[angle=-90]{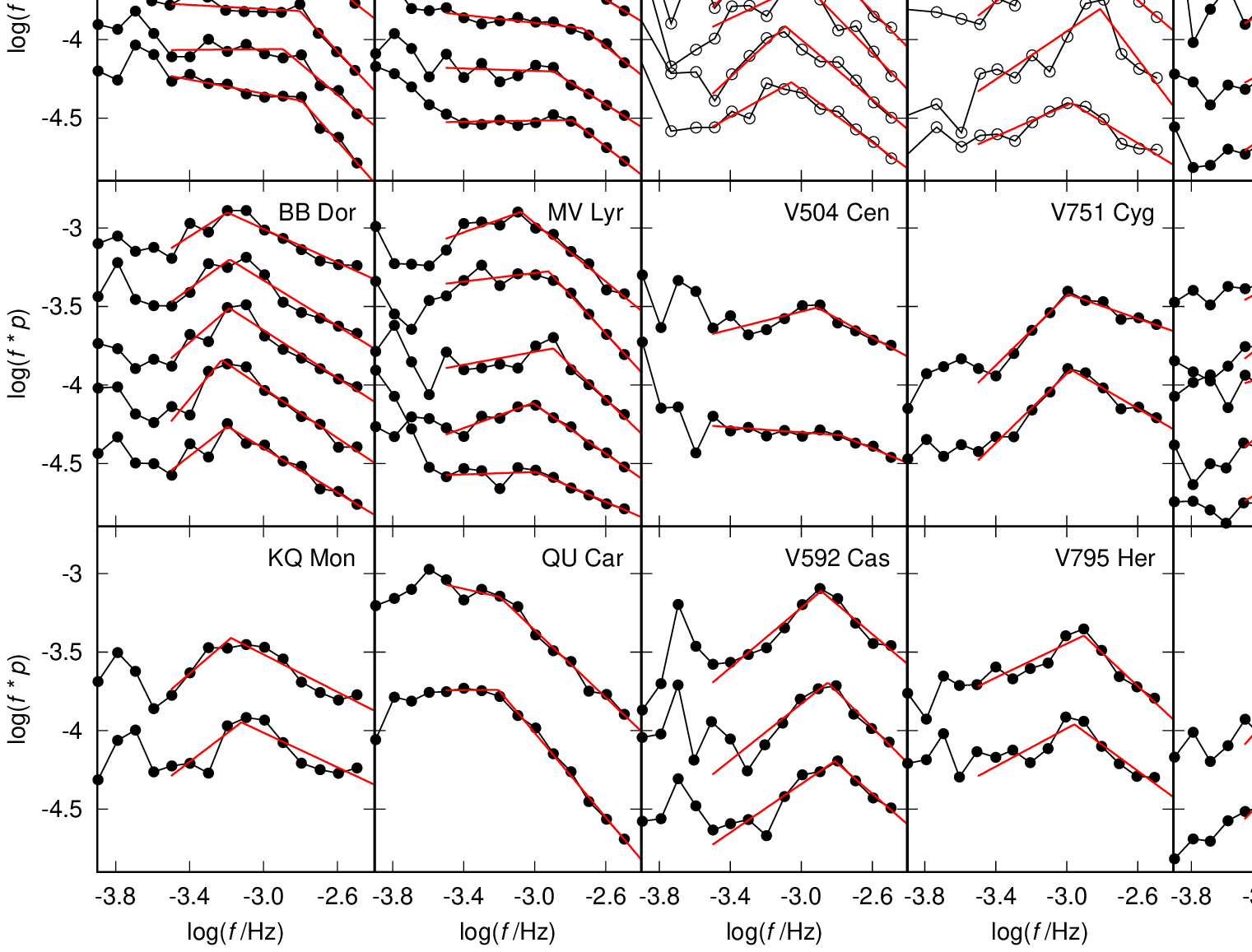}}
\caption{Examples of PDSs for individual systems with broken power law fits (red lines). Black points represent PDSs with $n_{\rm d} = 10$ and black circles are for $n_{\rm d} = 5$. Individual PDSs are offset vertically for better visualisation.}
\label{pdss}
\end{figure*}

\begin{figure}
\resizebox{\hsize}{!}{\includegraphics[angle=-90]{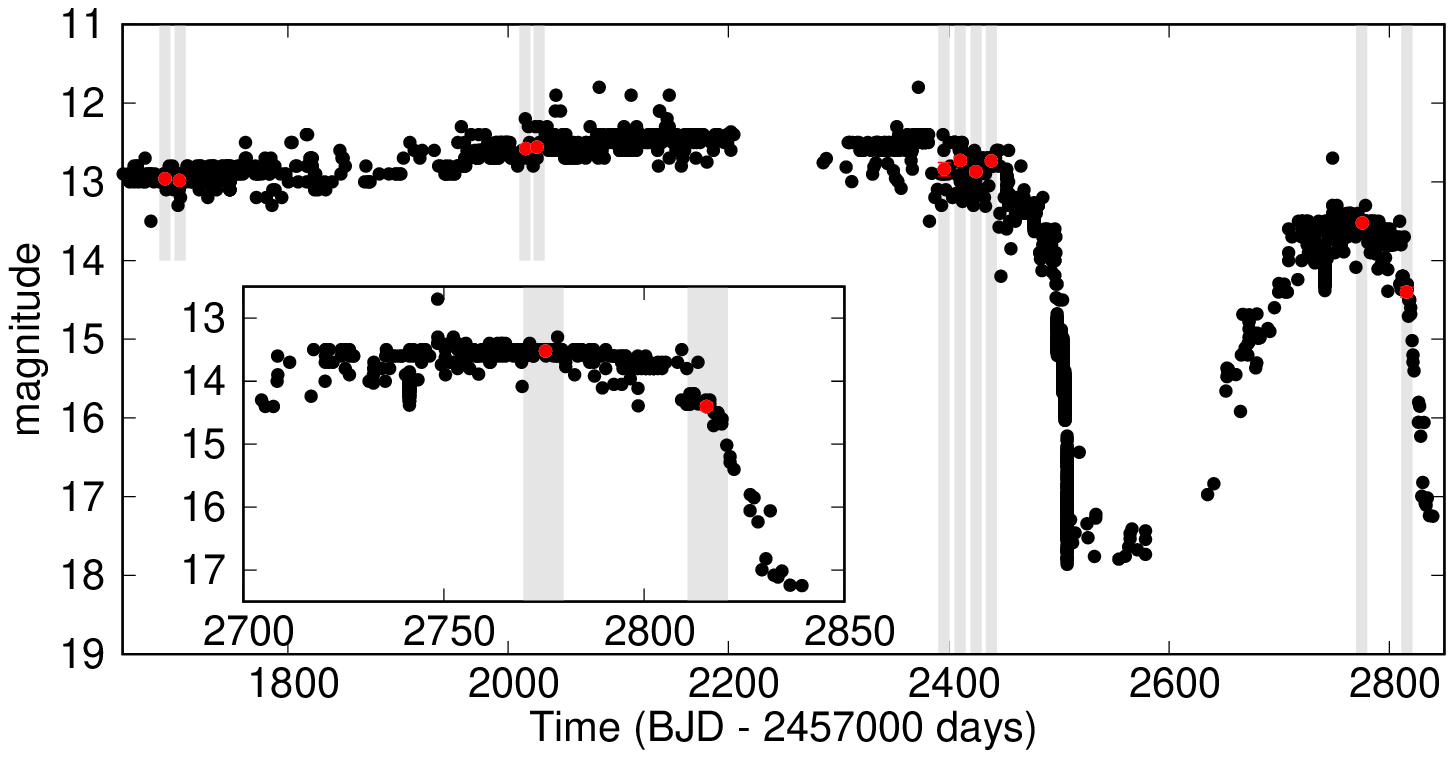}}
\caption{AAVSO light curve of MV\,Lyr with marked observation intervals of \tess\ by gray shaded areas. The inset panel shows a detail of the two most recent \tess\ observations falling before and at the beginning of the decline from the high optical state. Red points are mean brightness of the AAVSO data during the \tess\ observations.}
\label{lc_mvlyr}
\end{figure}

\begin{figure*}
\resizebox{\hsize}{!}{\includegraphics[angle=-90]{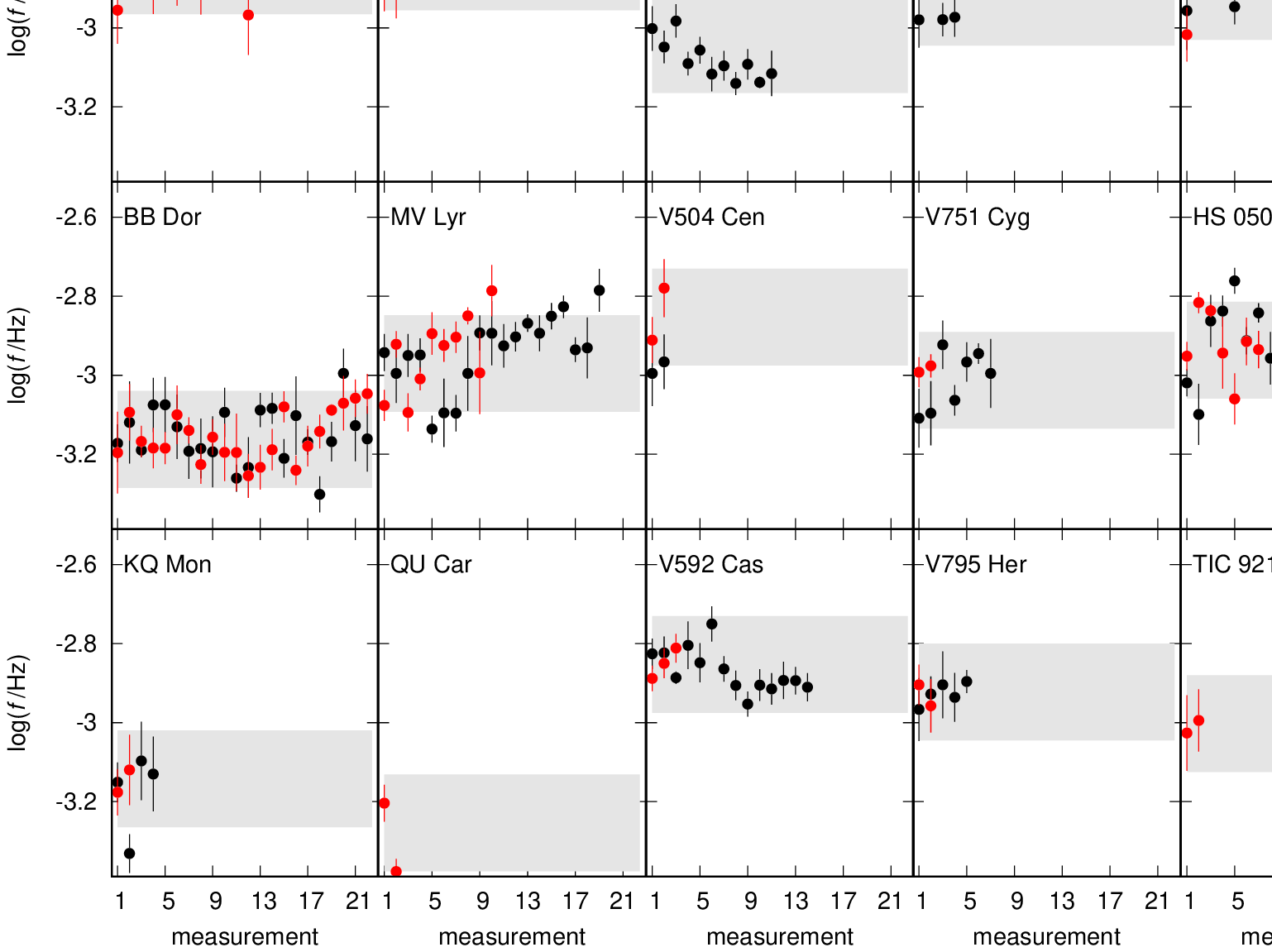}}
\caption{Measurements of $f_{\rm b}$ where conditions for positive detection were satisfied (see text for details). Red and black points represent the $n_{\rm d} = 10$ and 5 case, respectively. Grey shaded area is the natural scatter based on MV\,Lyr case. BB\,Dor case is not fully displayed for clarity, but the hidden measurements have the same scatter as shown (compared to gray shaded area).}
\label{pds_measurements}
\end{figure*}

Fig.~\ref{pdss} shows selected examples of PDSs with fits for objects where we found positive detections. Fig.~\ref{pds_measurements} shows measured $f_{\rm b}$ with errors for all positive detections. Both $n_{\rm d}$ of 5 and 10 are shown. As previously mentioned, the $n_{\rm d} = 5$ case is more scattered than $n_{\rm d} = 10$. Therefore, we rely on $n_{\rm d} = 10$ if possible. Only TT\,Ari and V704\,And does not have uninterrupted light curve portions long enough and only $n_{\rm d} = 5$ cases are shown. To judge whether the scatter of measurements is acceptable or not, we again used the well studied MV\,Lyr as a benchmark. The $n_{\rm d} = 10$ measurements scatter between values log($f$/Hz) = -2.79 and -3.09. All points are randomly redistributed around the value of log($f$/Hz) = -2.9, and only the last point (measurement 10) is slightly deviated toward higher frequencies. We investigated the long term AAVSO light curve and concluded that this measurement was made during a transition from a high to a low state (Fig.~\ref{lc_mvlyr}). \citet{dobrotka2020} showed that the $f_{\rm b}$ increases during such transition, therefore the deviation toward higher frequencies is natural. Since we need stationary data, we excluded this measurement from our analysis. The last but one observation is not made during the "standard" high state before day MJD = 2400, but apparently it occurred during a stable plateau. The brightness differs from that of all other observations, but since it is stabilised, the disc probably has all the physical characteristics of the "standard" high state. Finally, the measured $f_{\rm b}$ shows no suspicious deviation. All other \tess\ observations were taken during the "standard" high state. Therefore, as measurement scatter we take the frequency interval from log($f$/Hz) = -2.85 to -3.09 (gray shaded area). Based on more precise measurements using the superior spacecraft \kepler\, this scatter is natural and represents the well known $f_{\rm b}$ at approximately log($f$/Hz) = -3 ($f_2$ in Fig.~1 from \citealt{dobrotka2020}).

Let's see the Fig.~\ref{pds_measurements}. As a priority we take and describe the $n_{\rm d} = 10$ case. This is not possible only for TT\,Ari and V704\,And where we rely on $n_{\rm d} = 5$ data. All measurements lie within the gray shaded area except AH\,Men and already discussed MV\,Lyr. AH\,Men falls very well to the scatter interval too except one single point (11$^{\rm th}$ measurement). The AAVSO light curve is not covered enough to judge whether the system experienced similar brightness transition like MV\,Lyr. Anyhow, all other measurements are robust, therefore we excluded the deviated point from all subsequent analysis and discussions. All measured and selected $f_{\rm b}$ values displayed in Fig.~\ref{pds_measurements} are summarised in Tables~\ref{table_pds_prameters_10days} and \ref{table_pds_prameters_5days}.

\begin{figure}
\resizebox{\hsize}{!}{\includegraphics[angle=-90]{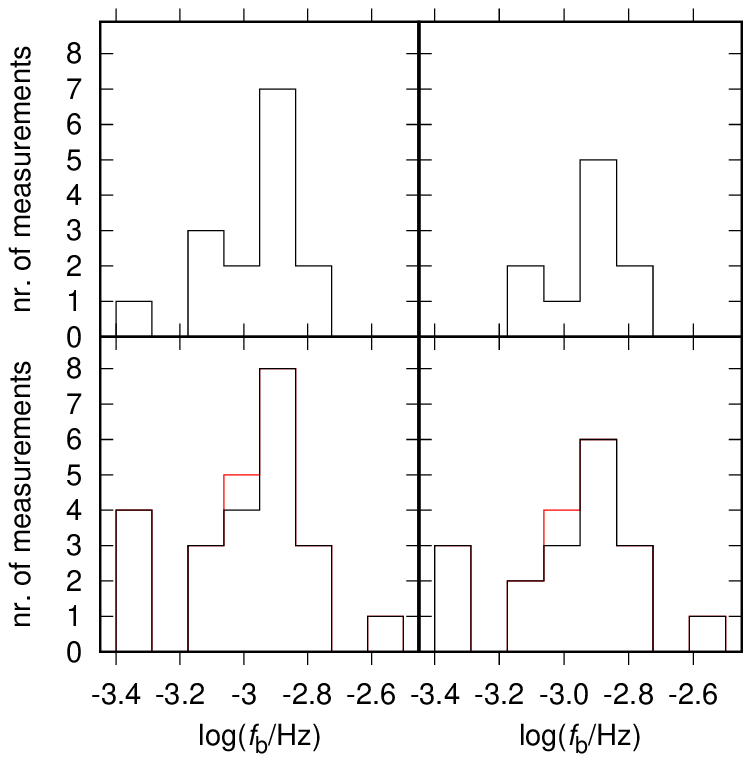}}
\caption{Histograms of weighted means (number of measurements) of $f_{\rm b}$ of individual systems. Left panels - systems with minimal number of selected $f_{\rm b}$ equal to two (AH\,Men, BB\,Dor, KQ\,Mon, KR\,Aur, MV\,Lyr, QU\,Car, TT\,Ari, V504\,Cen, V592\,Cas, V704\,And, V751\,Cyg, V795\,Her, V1193\,Ori, HS\,0506+7725, TIC\,92167387). Right panels - systems with minimal number of selected $f_{\rm b}$ equal to five (AH\,Men, BB\,Dor, KR\,Aur, MV\,Lyr, TT\,Ari, V592\,Cas, V751\,Cyg, V795\,Her, V1193\,Ori, HS\,0506+7725). Upper panels - measurements from this paper. Lower panels - same as upper panels but with added measurements from other works. Red line represents the uncertain V4743\,Sgr case.}
\label{histogram}
\end{figure}

To test the hypothesis from \citet{dobrotka2020} regarding the prevalence of $f_{\rm b}$ at approximately log($f_{\rm b}$/Hz) = -3, we calculated a histogram representing the statistical distribution of measured $f_{\rm b}$ values from log($f_{\rm b}$/Hz) = -3.4 to -2.5\footnote{Since the fit is performed on interval from log($f_{\rm b}$/Hz) = -3.5 to -2.4, the break frequency cannot fall to the boundaries.}. For this purpose we used weighted averages of measured $f_{\rm b}$ values using $n_{\rm d} = 10$ preferentially. This averages are summarised in Table~\ref{table_pds_prameters_systems}. The corresponding histograms with 8 bins\footnote{Comparable to Fig.~11 in \citet{dobrotka2020}.} are depicted in upper panels of Fig.~\ref{histogram}.
\begin{table}
\caption{Weighted averages of measured $f_{\rm b}$ for individual systems used for calculation of histograms in Fig.~\ref{histogram}.}
\begin{center}
\begin{tabular}{llrrr}
\hline
\hline
object & log($f_{\rm b}$/Hz) & $n_{\rm d}$ & $n_{\rm m}$ & $n_{\rm p}$\\
& & & & (\%)\\
\hline
AH\,Men		& $-2.813 \pm 0.017$ & 10 & 12 & 52\\
BB\,Dor		& $-3.145 \pm 0.008$ & 10 & 28 & 97\\
KQ\,Mon		& $-3.159 \pm 0.049$ & 10 & 2 & 100\\
KR\,Aur		& $-2.778 \pm 0.018$ & 10 & 6 & 100\\
MV\,Lyr		& $-2.937 \pm 0.012$ & 10 & 10 & 77\\
QU\,Car		& $-3.321 \pm 0.027$ & 10 & 2 & 50\\
TT\,Ari		& $-3.105 \pm 0.009$ & 5 & 11 & 100\\
V504\,Cen	& $-2.860 \pm 0.046$ & 10 & 2 & 100\\
V592\,Cas	& $-2.853 \pm 0.020$ & 10 & 3 & 100\\
			& $-2.881 \pm 0.009$ & 5 & 14 & 100\\
V704\,And	& $-2.939 \pm 0.026$ & 5 & 4 & 80\\
V751\,Cyg	& $-2.982 \pm 0.023$ & 10 & 2 & 100\\
			& $-2.989 \pm 0.017$ & 5 & 7 & 70\\
V795\,Her	& $-2.923 \pm 0.041$ & 10 & 2 & 100\\
			& $-2.913 \pm 0.021$ & 5 & 5 & 83\\
V1193\,Ori	& $-2.849 \pm 0.019$ & 10 & 3 & 75\\
			& $-2.852 \pm 0.013$ & 5 & 5 & 63\\
HS\,0506+7725	& $-2.866 \pm 0.012$ & 10 & 7 & 58\\
TIC\,92167387	& $-3.008 \pm 0.061$ & 10 & 2 & 100\\
\hline
\end{tabular}
\tablefoot{See text for details about the use of $n_{\rm d}$ and $n_{\rm m}$ for selection.}
\end{center}
\label{table_pds_prameters_systems}
\end{table}

First we show all systems from Fig.~\ref{pds_measurements} (upper left panel). This case is based on systems with minimal number of selected $f_{\rm b}$ measurements equal to two. Such number is too low to judge whether the $f_{\rm b}$ is stable or not. Larger criterion would be better to increase the confidence of the histogram. However, the larger this criterion, the lower the number of systems in the histogram. This is at the expense of the histogram resolution. Nevertheless, we show also such more conservative case with minimal number of $f_{\rm b}$ measurements equal to five. Focusing primary on the $n_{\rm d} = 10$ condition this excludes KQ\,Mon, QU\,Car, TT\,Ari, V504\,Cen, V592\,Cas, V704\,And, V751\,Cyg, V795\,Her, V1193\,Ori and TIC\,92167387. If $n_{\rm d} = 10$ case did not satisfy the condition of minimal detections of five, we used $n_{\rm d} = 5$ case instead. This brings back TT\,Ari, V592\,Cas, V751\,Cyg, V795\,Her and V1193\,Ori. The corresponding histogram is in the upper right panel of Fig.~\ref{histogram}.

\section{Discussion}

We investigated the optical flickering of CVs in high optical state. For this purpose we selected nova-like systems observed by \tess\ satellite. We searched for break frequency $f_{\rm b}$ in corresponding PDSs. The goal was to increase the low number (just five) of detections in \citet{dobrotka2020}. We found 13 new (15 in total) systems with positive detections of $f_{\rm b}$. Histograms in Fig.~\ref{histogram} depict the resulting statistics.

\subsection{PDS break frequency}

\citet{dobrotka2020} formulated a hypothesis that CVs in high state have a preferable $f_{\rm b}$ close to log($f$/Hz) = -3. The Fig.~\ref{histogram} supports this possibility. Apparently, the measurements cluster and culminate close to log($f$/Hz) = -3. The upper panels show the detections from this work, while the lower panels depict the statistics after including additional MV\,Lyr\footnote{For MV\,Lyr we added only values of log($f$/Hz) = -3.3 and -2.5, because $f_{\rm b}$ close to log($f$/Hz) = -3 was used from this work.}, KR\,Aur, V1504\,Cyg, V344\,Lyr and V4743\,Sgr values from Table~2 of \citet{dobrotka2020}. UU\,Aqr is not added because it is out of studied frequency range from log($f$/Hz) = -3.4 to -2.5.

The only problematic is V4743\,Sgr, identified as an intermediate polar with magnetic WD (\citealt{ness2003,kang2006}). Such magnetic CVs have truncated discs with the inner disc edge responsible for observed $f_{\rm b}$ very close to the log($f$/Hz) $\simeq$ -3 or higher (\citealt{revnivtsev2010,semena2014}). Usually, these $f_{\rm b}$ are close to the spin frequency of the WD. This is also the case of V4743\,Sgr (\citealt{dobrotka2021}). This means that the measured $f_{\rm b}$ can be of different origin than the studied flickering in non-magnetic CVs like in this work. However, all IPs have disc in the low state (\citealt{hameury2017}), while \citet{dobrotka2021} suggests that the disc in V4743\,Sgr is in a high state making this IP different from "standard" IPs. It is a post-nova. Therefore, it is possible that the measured $f_{\rm b}$ has different origin than in "standard" IPs, and it represents the studied flickering. But since this is only a possibility, we mark it as a red line in Fig.~\ref{histogram}.

The histogram in Fig.~\ref{histogram} is divided into 8 equally spaced bins with slightly narrower frequency extension than used in \citet{dobrotka2020}. This different spacing with slightly better resolution (0.113 vs. 0.144) allows to better identify the $f_{\rm b}$, which is between log($f$/Hz) = -2.95 and -2.84. Question is whether the observed dominant bin in the histogram represents a preferred or most frequent value, or it is just a random feature of a uniform distribution. \citet{dobrotka2021} simulated such a histogram with a uniform distribution and concluded that 31\% of simulations yield at least one bin with an equal or larger height than the dominant bin in their Fig.~11. This means that the statement that the dominant bin represents a preferred frequency has a confidence of only 69\%. Adding the V4743\,Sgr case this confidence increases to 91\%.

With the new histogram in Fig.~\ref{histogram}, we see that the V4743\,Sgr case do not fall into the dominant bin, and therefore does not increase the estimated confidence. We performed the same simulations as \citet{dobrotka2021}, and after one million repetitions, we got confidences of 99 and 96\% using only \tess\ detections from this work for a minimal number of two and five $f_{\rm b}$ measurements (upper panels of Fig.~\ref{histogram}), respectively. Adding all other measurements except V4743\,Sgr (lower panels of Fig.~\ref{histogram}) yield confidences of 96 and 85\%. Finally, adding also the questionable post-nova yields values of 95 and 81\%.

Apparently, the worst case is 81\%, but excluding the uncertain V4743\,Sgr it is 85\%. Therefore, we got a significant increase in confidence compared to the original 69\%, but still not enough for a definite conclusion, unless we want to rely on the less conservative case with a minimal number of $f_{\rm b}$ measurements of two. Nevertheless, we must be caution in adding measurements from different instruments. If \tess\ is able to detect only the dominant $f_{\rm b}$ in MV\,Lyr, and \kepler\ is able to see fainter PDS structures, we may discuss the dominant value close to log($f$/Hz) = -3, but it may be confusing to judge the statistical significance of the presence of other frequencies detected by \kepler\ only. Therefore, it is best to rely on measurements from this work based on \tess\ observations only. In such a case, the more conservative approach yield a confidence of 96\%. This is already high enough to conclude that nova-like CVs have very probably a preferred $f_{\rm b}$ between log($f$/Hz) = -2.95 and -2.84. Whether it is really a kind of preferred value or just a maximum in a smooth $f_{\rm b}$ distribution is not known yet due to the low number of $f_{\rm b}$ measurements.

\subsection{Dependance on physical parameters}

The next step is to understand the physical meaning of such a preferred value or distribution maximum. The best way is to search for a correlation of the $f_{\rm b}$ with basic physical parameters. Comparing $f_{\rm b}$ values with orbital periods did not yield any correlation. The data show just a scattered cloud of points. This is not surprising because the flickering in CVs and related objects is usually associated with the accretion disc and especially its inner regions (\citealt{bruch1992,bruch1996,zamanov1998,bruch2000,baptista2004,scaringi2014}), while the orbital period is more typically associated with the secondary star characteristics (\citealt{smith1998}).

The inner disc region potentially connected to the flickering is the inner disc edge. \citet{balman2012} suggests that the $f_{\rm b}$ represents the Keplerian frequency at this inner disc edge. If true, the $f_{\rm b}$ should be higher in the high state compared to the low state. However, opposite is observed in the case of nova-like MV\,Lyr (\citealt{dobrotka2020}). Nova-likes in the high state are similar to dwarf novae in outburst where the inner disc is build down to the WD surface or close to it. Once the CV changes states, the disc starts to be truncated and the Keplerian frequency should decrease during the transition. Opposite was observed in the transition of MV\,Lyr detected by the \kepler\ spacecraft. The dominant $f_{\rm b}$ at log($f$/Hz) = -3 increased together with another one or two adjacent PDS components. Anyhow, this does not tell us anything about the preference of any $f_{\rm b}$ value. Even if the $f_{\rm b}$ is generated by the inner disc edge, the preference would suggest some preferred inner disc radius. It is hard to test such a statement because measurements of the inner disc radii are rare.

An alternative interpretation is the sandwich model with a inner hot corona. Based on the accretion fluctuation propagation model an accretion inhomogeneity forms somewhere in the disc and propagates inwards. Such fluctuation follows the local physical conditions and local viscous time scale. Further away from the center is such fluctuation generated, larger is the corresponding time scale. If the fluctuations are generated in the inner corona, the outer radius of such geometrically thick disc determines the largest time scale of the fluctuations. Therefore, the larger the corona outer radius, the lower the $f_{\rm b}$ (Fig.~3 in \citealt{scaringi2014}). This may explain the increase of $f_{\rm b}$ during the transition of MV\,Lyr from high to low state. If $\dot{m}_{\rm acc}$ decreases during such a transition, the energy generation decreases too. Lower energy yields lower matter evaporation and the corona may shrink. As a consequence, this shrinking generates an increase of $f_{\rm b}$. Similar reasoning may be applied when considering mass of the central WD $m_{\rm WD}$. A larger $m_{\rm WD}$ generates a deeper potential well. This results in higher temperatures of the inner disc and the corona can be evaporated to further distances from the center. Such a more radially extended corona should generate a lower $f_{\rm b}$.

To test the $m_{\rm WD}$ scenario, we need to collect the masses measurements. In the case of nova-likes this is not easy because the accretion rate is too high to allow detection of a WD even at UV wavelengths. An accretion disc model or a combination of it with a WD model can be used instead. AH\,Men was studied using IUE spectroscopy (\citealt{gansicke1999}). Unfortunately, the determined $m_{\rm WD}$ is not unambiguous due to conflicts with distance constraints. The disc modelling yields values of 0.35 and 0.55\,M$_{\rm \odot}$. We use a mean value of $0.45 \pm 0.10$\,M$_{\rm \odot}$. \citet{godon2008} studied FUSE spectra of BB\,Dor and disc modelling derived a $m_{\rm WD}$ of 0.8\,M$_{\rm \odot}$. KQ\,Mon $m_{\rm WD}$ was determined by \citet{wolfe2013} using archival IUE spectra. Its value from disc modelling is estimated to 0.6\,M$_{\rm \odot}$. KR\,Aur has several $m_{\rm WD}$ estimates. \citet{shafter1983} used somewhat controversial method (as stated by the authors) using radial velocities and estimated value of 0.7\,M$_{\rm \odot}$ using spectra from ground observations. \citet{mizusawa2010} used IUE spectra and derived slightly lower mass of 0.6\,M$_{\rm \odot}$ using combination of disc and WD model. These two estimates agree well with another value of $0.59 \pm 0.17$\,M$_{\rm \odot}$ from \citet{ritter2003} catalogue. Most recently, \citet{rodriguezgil2020} used GTC spectra during low state, during which the WD is dominant and well seen. They derived quite a different mass from the previous estimates with a value of $0.94^{+0.15}_{-0.11}$\,M$_{\rm \odot}$. For MV\,Lyr \citet{hoard2004} derived a mass of $0.73 \pm 0.10$\,M$_{\rm \odot}$ from FUSE spectroscopy using a combined WD + disc model. The system was observed in a low state, therefore the spectrum was dominated by the WD features. $m_{\rm WD}$ of QU\,Car was determined using combined FUSE, HST and IUE spectra (\citealt{linnell2008}). Non-standard accretion disc model yields $m_{\rm WD}$ between 0.6 and 1.2\,M$_{\rm \odot}$. We use a mean value of $0.9 \pm 0.3$\,M$_{\rm \odot}$. $m_{\rm WD}$ in TT\,Ari can be estimated as 0.57\,M$_{\rm \odot}$ using spectra from ground observations and applying disc plus additional black body model (\citealt{belyakov2010}). \citet{huber1998} used photometric and spectroscopic data to study system parameters of V592\,Cas. The authors acknowledge that the method they used is criticised by several other authors\footnote{It is the same method as used for KR\,Aur by \citet{shafter1983} where the authors acknowledge the problematic aspect.}. Their estimated $m_{\rm WD}$ is $1.4^{+1.0}_{-0.6}$\,M$_{\rm \odot}$. We take 1.4\,M$_{\rm \odot}$ as an upper limit due to Chandrasekhar limit. Finally, WD of V795\,Her has a mass of approximately 0.8\,M$_{\rm \odot}$ by \citet{mizusawa2010} using IUE spectra and disc + WD atmosphere model. All values are summarised in Table.~\ref{table_system_prameters}.
\begin{table}
\caption{Used system parameters.}
\begin{center}
\begin{tabular}{lllr}
\hline
\hline
object & m$_{\rm WD}$ & $\dot{m}_{\rm acc}$ & $i$\\
& (M$_{\rm \odot}$) & (M$_{\rm \odot}$/yr) & (deg)\\
\hline
AH\,Men		& 0.35/0.55$^{\rm a}$ & $3 \times 10^{-10}$ $^{\rm a}$ & 20$^{\rm a}$\\
BB\,Dor		& 0.8$^{\rm b}$ & $10^{-9}$ $^{\rm b}$ & <10$^{\rm b}$\\
KQ\,Mon		& 0.6$^{\rm c}$ & $10^{-9}$ $^{\rm c}$ & 60 - 75$^{\rm c}$\\
KR\,Aur		& 0.7$^{\rm d}$ & & <40$^{\rm d}$\\
			& 0.6$^{\rm e}$ & $3 \times 10^{-10}$ $^{\rm e}$ & 41$^{\rm e}$\\
			& $0.59 \pm 0.17$ $^{\rm f}$ & & 38$^{\rm f}$\\
			& $0.94^{+0.15}_{-0.11}$ $^{\rm g}$ & & 47$^{\rm g}$\\
MV\,Lyr		& $0.73 \pm 0.10$ $^{\rm h}$ & $3 \times 10^{-9}$ $^{\rm i}$ & 12$^{\rm h}$\\
QU\,Car		& 0.6 - 1.2 $^{\rm j}$ & $10^{-7}$ - $10^{-6}$ $^{\rm j}$ & 40 - 60$^{\rm j}$\\
TT\,Ari		& 0.57$^{\rm k}$ & $4 \times 10^{-9}$ $^{\rm k}$ & 17 - 22.5$^{\rm k}$\\
V592\,Cas	& $1.4^{+1.0}_{-0.6}$ $^{\rm l}$ & & 28$^{\rm l}$\\
V795\,Her	& 0.8$^{\rm e}$ & $10^{-10}$ $^{\rm e}$ & 41$^{\rm e}$\\
\hline
AC\,Cnc		& & & 75.6$^{\rm f}$\\
DW\,UMa		& & & 82$^{\rm f}$\\
LX\,Ser		& & & 90$^{\rm f}$\\
RR\,Pic		& & & 65$^{\rm f}$\\
RW\,Tri		& & & 70.5$^{\rm f}$\\
TV\,Col		& & & 70$^{\rm f}$\\
UU\,Aqr		& & & 78$^{\rm f}$\\
UX\,UMa		& & & 70$^{\rm f}$\\
V348\,Pup	& & & 81.1$^{\rm f}$\\
\hline
\end{tabular}
\tablefoot{First part is for systems with $f_{\rm b}$ detection, while second part is only for systems with red noise PDSs.}
\end{center}
\tablebib{
$^{\rm a}$\citet{gansicke1999}, $^{\rm b}$\citet{godon2008}, $^{\rm c}$\citet{wolfe2013}, $^{\rm d}$\citet{shafter1983}, $^{\rm e}$\citet{mizusawa2010}, $^{\rm f}$\citet{ritter2003}, $^{\rm g}$\citet{rodriguezgil2020}, $^{\rm h}$\citet{hoard2004}, $^{\rm i}$\citet{linnell2005}, $^{\rm j}$\citet{linnell2008}, $^{\rm k}$\citet{belyakov2010}, $^{\rm l}$\citet{huber1998}
}
\label{table_system_prameters}
\end{table}

\begin{figure}
\resizebox{\hsize}{!}{\includegraphics[angle=-90]{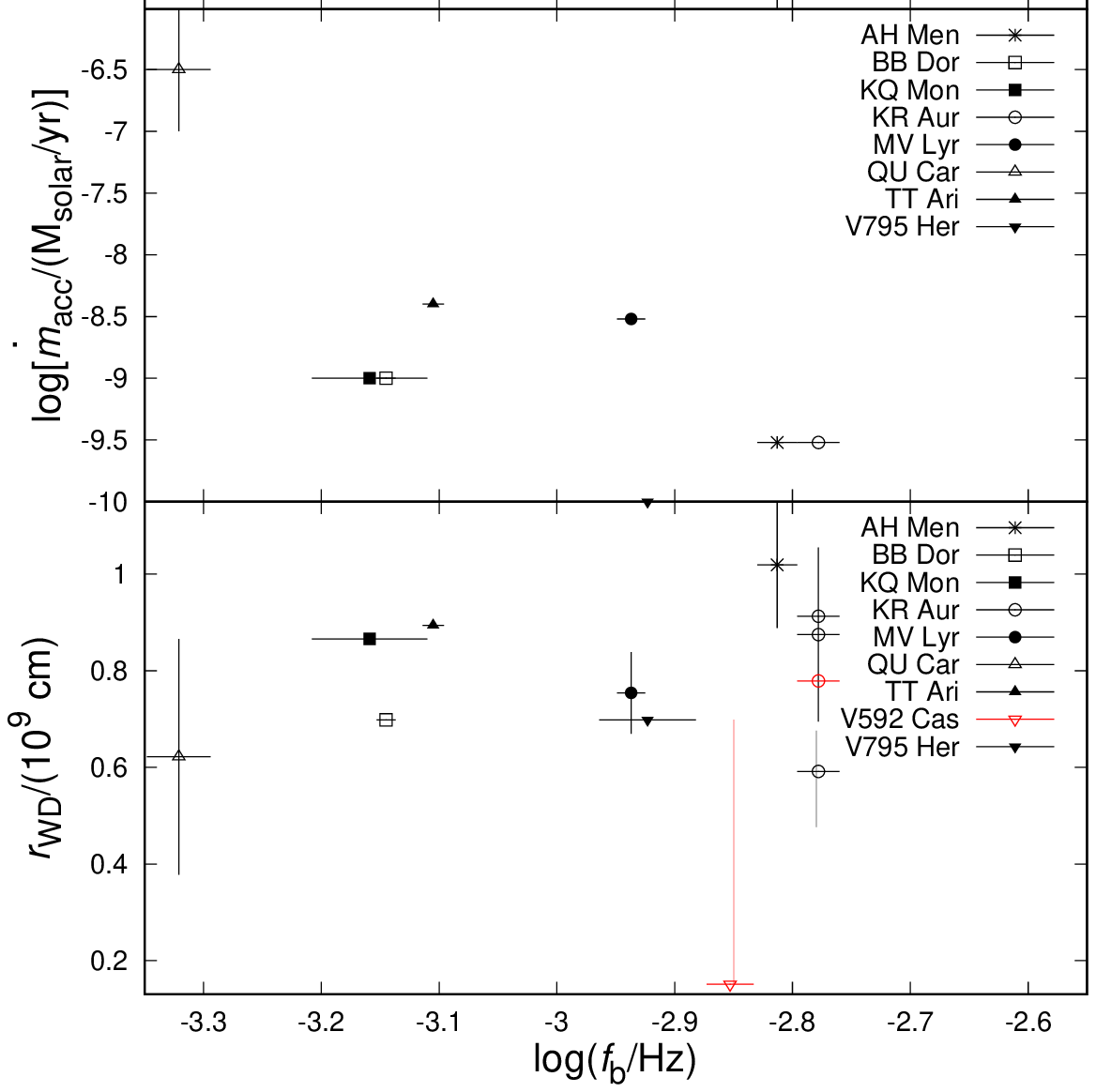}}
\caption{$m_{\rm WD}$ (upper panel), $\dot{m}_{\rm acc}$ (middle panel) and $r_{\rm WD}$ (bottom panel) vs $f_{\rm b}$ from Table~\ref{table_pds_prameters_systems}. Red color represents uncertain measurements.}
\label{wdmass_frekv}
\end{figure}

Upper panel of Fig.~\ref{wdmass_frekv} shows the correlation between $m_{\rm WD}$ and $f_{\rm b}$ of the mentioned systems. The first view is rather uncertain. Ignoring the problematic measurements shown by red color, mainly the V592\,Cas case, the situation is clearer. Only the too scattered KR\,Aur complicates the judgement. Without the highest $m_{\rm WD}$ estimate there is a tendency to see the anticipated correlation. $f_{\rm b}$ decreases with increasing $m_{\rm WD}$. The scatter of points is relatively large, but this is natural because determination of $m_{\rm WD}$ is very uncertain and is usually done with large errors. Therefore, we cannot definitely conclude whether the correlation is real or if the redistribution of points is just random. There are three ways of answering whether the correlation is real. We should either refine or revise $m_{\rm WD}$ of KR\,Aur and QU\,Car, increase the number of measured $f_{\rm b}$ for systems with known $m_{\rm WD}$, or to get $f_{\rm b}$ measurements for CVs with high $m_{\rm WD}$ (above 1\,M$_{\rm \odot}$) to populate upper right region of upper panel of Fig.~\ref{wdmass_frekv}.

If $f_{\rm b}$ is really correlated with $m_{\rm WD}$, the preferred value between log($f$/Hz) = -2.95 and -2.84 suggests existence of a preferred $m_{\rm WD}$. \citet{zorotovic2011} studied the $m_{\rm WD}$ distribution in CVs and their Fig.~7 shows a maximum at 0.8\,M$_{\rm \odot}$. This can represent the mentioned preferred value, and in such case it would be just a maximum of an othervise smooth distribution. Best way to check whether the maximum in $f_{\rm b}$ represents the observed maximum in $m_{\rm WD}$ is to fit the data in upper panel of Fig.~\ref{wdmass_frekv} with a linear function. Unfortunately, the fits have too large errors mainly for the first coefficient defining correlation or anticorrelation\footnote{Using all black data we got $m_{\rm WD} = (-0.07 \pm 0.84) {\rm log}(f_{\rm b}) + (-0.26 \pm 0.28)$\,M$_{\rm \odot}$, while excluding the KR\,Aur case because of too scattered mass estimate we got $m_{\rm WD} = (-0.81 \pm 1.05) {\rm log}(f_{\rm b}) + (-0.49 \pm 0.34)$\,M$_{\rm \odot}$. We did not use $m_{\rm WD}$ errors since they are not known for every system. Adding at least 0.1\,M$_{\rm \odot}$ as error, the fit is even worse, i.e. the relative errors are larger or the direction get positive in the case of all black data.}. Moreover, \citet{dobrotka2020} mentioned that three possible groups of $f_{\rm b}$ could be present, but due to very low number of measurements it is not possible to confirm or reject the statement. \citet{wijnen2015} theoretically studied $m_{\rm WD}$ distribution and they found three local maxima in simulated histograms. Therefore, a possibility that $f_{\rm b}$ distribution has more complicated structure has physical meaning. However, the histogram in \citet{dobrotka2020} is constructed using well studied MV\,Lyr, V1504\,Cyg and V344\,Lyr cases where multi-components PDSs were found. Therefore, such histogram is contaminated by multiple measurements for one single $m_{\rm WD}$. In such case the multiple structure of the histogram rather represents different sources of the fast variability with different $f_{\rm b}$, and the potential correlation between $m_{\rm WD}$ and $f_{\rm b}$ would be for only some specific $f_{\rm b}$. MV\,Lyr and V1504\,Cyg PDS show one dominant break and all other features are relatively weak. This dominant $f_{\rm b}$ seen in \tess\ data can be correlated with $m_{\rm WD}$.

$m_{\rm WD}$ and its deeper potential well is not the only reason for higher temperature of the disc. Larger increased energy generation and subsequent stronger evaporation of matter into the corona can be produced by higher $\dot{m}_{\rm acc}$ too. Middle panel of Fig.~\ref{wdmass_frekv} shows this case. We used values from Table.~\ref{table_system_prameters}. It seems that $f_{\rm b}$ decreases with increasing $\dot{m}_{\rm acc}$. However, this correlation is based mainly on one single deviated point representing QU\,Car. Otherwise, the points does not show any significant correlation. Finally, both the $m_{\rm WD}$ and $\dot{m}_{\rm acc}$ together can play a role. Higher potential well of a massive WD and higher $\dot{m}_{\rm acc}$ can generate larger radius of an evaporated corona. For a detailed investigation much more systems needs to be studied.

The confirmation of corona scenario via such $f_{\rm b}$ vs system parameters correlation has very important consequence. As mentioned in the introduction, the ratio of X-ray to optical luminosity is of the order of 0.1 (\citealt{dobrotka2020}) or even 0.01 - 0.001 (\citealt{balman2014}). How is it then possible that X-rays are able to generate the observed optical radiation? \citet{scaringi2014} proposed reprocessing which apparently cannot work. An alternative must be searched elsewhere. The variability can be generated by the corona. The reprocessing of X-rays from the corona generates only faint response. Such response can have a form of small side-lobes found in the shot profile by \citet{dobrotka2019}. The shot profile has another and dominant feature, the central spike. This has too large amplitude to be explainable by the reprocessing. It must originate directly from the central geometrically thin disc. The variability in the corona is generated by propagation of accretion fluctuations (\citealt{scaringi2014}). If these fluctuations somehow influence the underlying geometrically thin disc flow, the time scale of the optical flickering can have the same time scales. Such influence can be re-condensation of the corona back into the thin disc (\citealt{meyer2007}). If the corona generates the log($f$/Hz) = -3 variability via mass accretion fluctuations, these fluctuations re-condensate and propagate in the thin disc. This can generate optical variability with the same frequency.

Finally, for authors identifying boundary layer as the source of the flickering it is important to relate $f_{\rm b}$ with the WD radius $r_{\rm WD}$. We display such case in bottom panel of Fig.~\ref{wdmass_frekv}. For WD radius we used relation by \citet{nauenberg1972}, hence the figure looks similar to the $m_{\rm WD}$ vs $f_{\rm b}$ case. Since the relation between $m_{\rm WD}$ and $r_{\rm WD}$ is not linear, the increase of $r_{\rm WD}$ with increasing $f_{\rm b}$ is accentuated yielding better linear fits than in the $m_{\rm WD}$ case. While the fits of $m_{\rm WD}$ vs $f_{\rm b}$ did not yield definite slope character (rising or declining) due to large parameter errors, the $r_{\rm WD}$ case is much better with clearly positive slope\footnote{Using all black data we got $r_{\rm WD} = (1.47 \pm 0.71) {\rm log}(f_{\rm b}) + (0.23 \pm 0.24)\,10^9$\,cm, while excluding the KR\,Aur case because of too scattered mass estimate we got $r_{\rm WD} = (2.12 \pm 0.91) {\rm log}(f_{\rm b}) + (0.43 \pm 0.30)\,10^9$\,cm. Like in $m_{\rm WD}$ case we did not use errors because they are not known for every system, therefore the fit is just rough and orientational.}. However, the derivation of $r_{\rm WD}$ is not independent from the derivation of $m_{\rm WD}$, and the "better" fits are just results of the non-linear transformation. Therefore, we can not conclude any fit "superiority", and the character of data is still uncertain and another measurements are needed as concluded for $m_{\rm WD}$.

\subsection{Detection vs non-detection of $f_{\rm b}$}

An indirect indication that the central disc region is the source of the studied flickering can be get from the discussion about detection vs non-detection of $f_{\rm b}$. First idea is that the detection depends on the brightness of the binary. This is valid for systems with dominant white noise. Those systems have the lowest mean fluxes, therefore Poisson noise dominates the corresponding PDSs. However, all other systems with $f_{\rm b}$ detection or PDS dominated by red noise are equally distributed with no preferred fluxes except bright TT\,Ari and QU\,Car. These two systems have significantly higher fluxes, and $f_{\rm b}$ detection. But only two systems with specific flux does not say anything about any flux preference.

Apparently, the reason of detection vs non-detection must be searched elsewhere. The localisation of the flickering source can be the clue. If the source is the inner geometrically thin disc, either radiating directly or reprocessing X-rays from the central corona, it must be seen. This can be problematic in eclipsing systems or in systems with high inclination. In such conditions the central disc can be obscured by the outer disc edge. Investigating the inclinations (Table.~\ref{table_system_prameters}) of systems with detected $f_{\rm b}$ we see that all have values below 75$^{\circ}$. Taking the lower estimate of KQ\,Mon the inclinations are below 60$^{\circ}$. On the other hand, systems with red noise PDSs are eclipsing like AY Psc (\citealt{szkody1989}), BH\,Lyn (\citealt{andronov1986,richter1989,andronov1989}), NSV\,1907 (\citealt{hummerich2017}), or have higher inclinations (Table.~\ref{table_system_prameters}). Clearly, the inclinations in red noise systems is higher than 65$^{\circ}$. Therefore, there is a possible overlap of inclinations only in the cases of KQ\,Mon with the low end of red noise systems. Othervise, the inclinations are very different implying that $f_{\rm b}$ is seen only in low inclination systems.

Another reason of missing $f_{\rm b}$ also related to the inner disc regions is the IP nature of the system. In these CVs the central disc is truncated, therefore missing, due to magnetic field of the WD. This is the case of for example RX\,J2133.7+5107 (\citealt{bonnet2006}) where we see only red noise with clear spin frequency seen as one significantly deviated PDS bin. The PDS is not deformed like examples in the left panel of Fig.~\ref{pdss_bad}, therefore the red noise character is clear. V533\,Her is a similar case with no $f_{\rm b}$ detection and being an IP candidate (\citealt{worpel2020}). The PDS show ambiguous shape with not always present (variable from observation to observation) PDS structure near potential spin frequency of log($f$/Hz) = -3.15 (\citealt{rodriguez2002}). We should exclude this binary from analysis like we did for examples in the left panel of Fig.~\ref{pdss_bad}, but since the spin frequency is not obvious we kept it. Anyhow, we did not get any $f_{\rm b}$ detection, therefore this case is also consistent with missing inner disc region in IPs.

Finally, worth to discuss is CP\,Pup case. The PDS is neither pure red noise neither a white noise. Red noise is clear up to approximately log($f$/Hz) = -2.7 to -3, and a white noise plateau continues for higher frequencies. \citet{szkody1988} reported an inclination of 32$^{\circ}$ or 37$^{\circ}$. Based on our inclination assumption, such system should show a clear $f_{\rm b}$. However, the authors reported a potential WD mass of only 0.18\,M$_{\rm \odot}$ for mass ratio of 1. Similar potential low mass of 0.12\,M$_{\rm \odot}$ or 0.27\,M$_{\rm \odot}$ concluded also \citet{duerbeck1987}. In the case of existing relation between $f_{\rm b}$ and $m_{\rm WD}$ and based on the characteristics of Fig.~\ref{wdmass_frekv} the potential $f_{\rm b}$ (if present) can be outside of the studied frequency range for such low $m_{\rm WD}$ or can be hidden in the high frequency part of the PDS which is dominated by Poisson white noise. Therefore, CP\,Pup does not contradict the low inclination criterion for $f_{\rm b}$ detection, and thanks to its probable low $m_{\rm WD}$ it is a very suitable object for testing any relation between $f_{\rm b}$ and $m_{\rm WD}$.

\section{Summary and conclusions}

\citet{dobrotka2020} sumarised characteristic $f_{\rm b}$ of the flickering in CVs. The authors found that it is very probable that PDSs of these systems have a preferred value close to log($f$/Hz) = -3 but only in high optical state. However, the probability of such a conclusion is only 69\% (\citealt{dobrotka2021}). It is still possible that the resulting histogram of measured $f_{\rm b}$ is the result of a random distribution. More measurements are needed to improve the statistics.

We selected nova-like CVs observed by \tess\ and we searched for $f_{\rm b}$ in PDSs. These systems are the majority of their life-time in a high optical state. We selected uninterrupted light curve portions with durations of 5 and 10 days. We divided these portions into ten equally spaced subsamples and we calculated the mean PDSs. We focused our search on the frequency interval from log($f$/Hz) = -3.5 to -2.4 looking for the mentioned mHz $f_{\rm b}$.

We found 15 positive detections, and the resulting histogram shows the searched preferred value (maximum) close to the log($f$/Hz) = -3. Thanks to the higher number of measurements compared to \citet{dobrotka2020} we refined this value to the interval from log($f$/Hz) = -2.95 to -2.84. The probability that this histogram maximum is not the result of just a random distribution is 96\% in more conservative case. Apparently, this is much larger than previous value of 69\%. Moreover, it is possible that the $f_{\rm b}$ is correlated with WD mass; the higher the mass, the lower the $f_{\rm b}$. However, such a statement has only low confidence due to the still low number of studied systems with known WD mass and needs to be confirmed by further work. We need to increase the number of systems where the $f_{\rm b}$ is detected, and/or we should focus on systems with lower values close to log($f$/Hz) = -3.5 to populate empty region of the WD mass vs $f_{\rm b}$ parametric space.

As mentioned only 15 systems show clear $f_{\rm b}$ detection. All other objects have PDS dominated by white or red noise. While the white noise is the result of low brightness, the red noise is seen only in high inclination systems.

Both the inclination dependance and possible correlation between $f_{\rm b}$ and WD mass points to inner disc regions as the source of the detected $f_{\rm b}$. This region is not well seen in high inclination systems or completely obscured in eclipsing CVs. The potential correlation between $f_{\rm b}$ and WD mass could be explained by inner hot corona as a source of $f_{\rm b}$ which is radially more extended for more massive WDs.

\section*{Acknowledgement}

This work was supported by the Slovak grant VEGA 1/0576/24. We acknowledge with thanks the variable star observations from the AAVSO International Database contributed by observers worldwide and used in this research.

\bibliographystyle{aa}
\bibliography{mybib}

\label{lastpage}

\begin{appendix}
\section{Additional tables}

Tables~\ref{table_pds_prameters_10days} and \ref{table_pds_prameters_5days} show all selected $f_{\rm b}$ values for $n_{\rm d} = 10$ and $n_{\rm d} = 5$, respectively.
\begin{table*}
\caption{Selected $f_{\rm b}$ for systems with positive detection using $n_{\rm d} = 10$.}
\begin{center}
\begin{tabular}{lcccccccc}
\hline
\hline
object & start & log($f_{\rm b}$/Hz) & start & log($f_{\rm b}$/Hz) & start & log($f_{\rm b}$/Hz) & start & log($f_{\rm b}$/Hz)\\
& (MJD) & & (MJD) & & (MJD) & & (MJD)\\
\hline
AH\,Men			& 1325.297 & -2.955(085) & 1339.656 & -2.773(051) & 1424.554 & -2.804(037) & 1599.312 & -2.896(069)\\
				& 1613.758 & -2.795(034) & 2144.514 & -2.882(062) & 2158.861 & -2.753(055) & 2144.514 & -2.890(077)\\
				& 2158.861 & -2.784(072) & 2335.082 & -2.724(066) & 2335.078 & -2.612(031) & 2361.773 & -2.967(101)\\
BB\,Dor			& 1327.816 & -3.196(103) & 1354.108 & -3.094(072) & 1368.601 & -3.168(040) & 1424.556 & -3.184(051)\\
				& 1438.260 & -3.185(041) & 1451.620 & -3.100(074) & 1478.431 & -3.140(033) & 1491.903 & -3.226(049)\\
				& 1505.013 & -3.157(045) & 1518.378 & -3.196(073) & 1545.205 & -3.196(098) & 2036.278 & -3.255(056)\\
				& 2049.153 & -3.233(056) & 2075.156 & -3.189(052) & 2115.886 & -3.080(039) & 2130.210 & -3.241(037)\\
				& 2144.516 & -3.180(051) & 2158.863 & -3.143(043) & 2174.230 & -3.089(015) & 2187.239 & -3.071(069)\\
				& 2201.734 & -3.058(047) & 2215.435 & -3.047(050) & 2229.231 & -3.180(038) & 2242.585 & -3.125(068)\\
				& 2282.712 & -3.142(036) & 2295.900 & -3.242(059) & 2361.771 & -3.250(032) & 2375.861 & -3.174(087)\\
KQ\,Mon			& 2229.049 & -3.177(059) & 2242.403 & -3.120(089) & & & &\\
KR\,Aur			& 2474.172 & -2.878(081) & 2487.182 & -2.894(081) & 2500.394 & -2.728(034) & 2513.920 & -2.796(070)\\
				& 2526.052 & -2.788(029) & 2540.164 & -2.767(040) & & & &\\
MV\,Lyr			& 1683.356 & -3.077(039) & 1697.347 & -2.922(034) & 2010.269 & -3.094(048) & 2023.117 & -3.009(029)\\
				& 2390.656 & -2.895(054) & 2405.332 & -2.925(041) & 2419.992 & -2.904(039) & 2433.721 & -2.850(021)\\
				& 2769.902 & -2.994(104) & 2810.874 & -2.786(065) & & & & \\
QU\,Car			& 2321.031 & -3.204(047) & 2347.302 & -3.377(032) & & & &\\
V504\,Cen		& 2334.205 & -2.911(058) & 2347.305 & -2.779(073) & & & &\\
V592\,Cas		& 1777.734 & -2.888(032) & 1957.269 & -2.850(038) & 1970.969 & -2.811(036) & &\\
V751\,Cyg		& 2797.103 & -2.992(038) & 2810.874 & -2.976(029) & & & &\\
V795\,Her		& 2010.271 & -2.904(051) & 2023.118 & -2.957(068) & & & &\\
V1193\,Ori		& 1451.561 & -3.017(068) & 2174.234 & -2.809(028) & 2187.201 & -2.862(029) & &\\
HS\,0506+7725	& 1816.088 & -2.952(036) & 1828.970 & -2.816(027) & 1842.507 & -2.837(017) & 1856.400 & -2.944(090)\\
				& 1983.630 & -3.060(065) & 1996.914 & -2.915(060) & 2010.264 & -2.935(047) & &\\
TIC\,92167387	& 2309.017 & -3.026(096) & 2322.400 & -2.995(079) & & & &\\
\hline
\end{tabular}
\tablefoot{Start denotes the starting time of the selected light curve portion with MJD = JD - 2457000. The values in parenthesis represent the errors.}
\end{center}
\label{table_pds_prameters_10days}
\end{table*}
\begin{table*}
\caption{The same as Table~\ref{table_pds_prameters_10days} but for $n_{\rm d} = 5$.}
\begin{center}
\begin{tabular}{lcccccccc}
\hline
\hline
object & start & log($f_{\rm b}$/Hz) & start & log($f_{\rm b}$/Hz) & start & log($f_{\rm b}$/Hz) & start & log($f_{\rm b}$/Hz)\\
& (MJD) & & (MJD) & & (MJD) & & (MJD)\\
\hline
BB\,Dor			& 1332.816 & -3.172(048) & 1339.655 & -3.119(104) & 1368.601 & -3.190(018) & 1396.641 & -3.075(069)\\
				& 1410.903 & -3.075(070) & 1424.556 & -3.131(082) & 1429.556 & -3.193(070) & 1438.260 & -3.186(076)\\
				& 1443.260 & -3.194(090) & 1451.620 & -3.094(062) & 1468.273 & -3.261(034) & 1478.431 & -3.234(077)\\
				& 1483.432 & -3.088(043) & 1496.905 & -3.084(040) & 1505.013 & -3.211(049) & 1510.014 & -3.102(099)\\
				& 1518.378 & -3.170(032) & 1534.998 & -3.302(046) & 1558.541 & -3.168(050) & 1573.258 & -2.996(062)\\
				& 1587.151 & -3.128(090) & 1673.056 & -3.161(083) & 2049.153 & -3.173(074) & 2061.851 & -3.179(036)\\
				& 2080.156 & -3.211(077) & 2088.240 & -3.166(095) & 2102.328 & -3.091(071) & 2115.886 & -3.195(069)\\
				& 2120.886 & -3.080(072) & 2130.210 & -3.020(102) & 2135.210 & -3.257(080) & 2144.516 & -3.146(034)\\
				& 2149.516 & -3.134(044) & 2158.863 & -3.195(072) & 2163.863 & -3.046(064) & 2174.230 & -3.145(084)\\
				& 2179.230 & -3.191(042) & 2201.734 & -2.993(065) & 2206.735 & -3.067(043) & 2215.435 & -2.967(082)\\
				& 2220.436 & -3.245(050) & 2234.232 & -3.195(084) & 2282.712 & -3.060(053) & 2287.713 & -3.152(036)\\
				& 2295.900 & -3.248(047) & 2300.902 & -3.170(090) & 2323.765 & -3.136(077) & 2337.353 & -3.065(058)\\
				& 2351.849 & -3.166(064) & 2361.771 & -3.383(035) & 2366.771 & -3.168(055) & 2375.861 & -3.183(086)\\
				& 2964.040 & -3.054(058) & 2969.550 & -3.096(070) & 2982.406 & -3.127(054) & 2989.842 & -3.149(032)\\
				& 2995.098 & -3.083(042) & 3007.750 & -3.096(065) & 3015.381 & -2.948(065) & 3021.403 & -3.093(075)\\
				& 3028.757 & -3.085(062) & 3034.306 & -3.049(078) & 3041.114 & -3.075(070) & 3083.690 & -3.096(094)\\
				& 3112.190 & -3.179(063) & 3126.641 & -3.095(093) & 3155.045 & -3.326(039) & 3168.617 & -3.162(075)\\
				& 3182.502 & -3.064(064) & & & & & &\\
KQ\,Mon			& 2229.049 & -3.151(050) & 2234.049 & -3.332(049) & 2242.403 & -3.097(099) & 2247.406 & -3.130(095)\\
MV\,Lyr			& 1683.356 & -2.942(047) & 1688.358 & -2.995(075) & 1697.347 & -2.950(054) & 1702.348 & -2.949(042)\\
				& 2010.269 & -3.137(034) & 2015.270 & -3.095(087) & 2023.117 & -3.096(046) & 2028.117 & -2.996(095)\\
				& 2390.656 & -2.893(044) & 2395.656 & -2.894(081) & 2405.332 & -2.926(055) & 2410.332 & -2.903(037)\\
				& 2419.992 & -2.868(022) & 2424.993 & -2.894(045) & 2433.721 & -2.850(033) & 2438.723 & -2.827(029)\\
				& 2774.902 & -2.935(031) & 2783.534 & -2.931(077) & 2810.874 & -2.785(054) & &\\
TT\,Ari			& 2447.693 & -3.002(056) & 2461.243 & -3.048(041) & 2474.183 & -2.982(042) & 2487.186 & -3.090(030)\\
				& 2492.187 & -3.057(034) & 3208.791 & -3.117(044) & 3221.701 & -3.096(037) & 3227.661 & -3.141(029)\\
				& 3235.439 & -3.092(038) & 3240.714 & -3.138(015) & 3253.592 & -3.116(057) & &\\
V504\,Cen		& 2334.205 & -2.995(082) & 2339.205 & -2.966(069) & & & &\\
V592\,Cas		& 1764.688 & -2.826(039) & 1777.734 & -2.823(042) & 1782.734 & -2.886(016) & 1957.269 & -2.804(060)\\
				& 1962.271 & -2.848(049) & 1970.969 & -2.750(045) & 1975.969 & -2.864(032) & 2853.358 & -2.906(038)\\
				& 2867.465 & -2.953(032) & 2874.823 & -2.905(040) & 2883.011 & -2.915(040) & 2889.639 & -2.893(047)\\
				& 2897.224 & -2.894(035) & 2903.600 & -2.910(036) & & & &\\
V704\,And		& 1777.735 & -2.979(071) & 2853.359 & -2.812(053) & 2867.466 & -2.979(042) & 2874.823 & -2.973(049)\\
V751\,Cyg		& 1711.368 & -3.109(074) & 1724.944 & -3.096(081) & 2797.103 & -2.923(061) & 2802.104 & -3.064(038)\\
				& 2810.874 & -2.967(050) & 2815.874 & -2.945(026) & 2831.944 & -2.995(087) & &\\
V795\,Her		& 1988.187 & -2.967(080) & 2001.467 & -2.928(044) & 2010.271 & -2.904(085) & 2015.272 & -2.936(062)\\
				& 2028.119 & -2.896(029) & & & & & &\\
V1193\,Ori		& 1456.561 & -2.956(100) & 2174.234 & -2.752(033) & 2179.234 & -2.832(020) & 2187.201 & -2.902(025)\\
				& 2192.201 & -2.946(045) & & & & & &\\
HS\,0506+7725	& 1816.088 & -3.019(034) & 1821.088 & -3.099(077) & 1828.970 & -2.863(066) & 1833.970 & -2.837(039)\\
				& 1842.507 & -2.761(033) & 1847.509 & -2.913(034) & 1856.400 & -2.842(025) & 1983.630 & -2.957(067)\\
				& 1988.633 & -2.975(068) & 1996.914 & -2.977(062) & 2001.916 & -2.845(047) & 2015.266 & -2.926(025)\\
				& 2023.111 & -2.797(064) & 3292.486 & -3.096(073) & 3306.197 & -2.957(081) & &\\
\hline
\end{tabular}
\end{center}
\label{table_pds_prameters_5days}
\end{table*}

\end{appendix}

\end{document}